\documentclass[pra,superscriptaddress,10pt]{revtex4}
\usepackage{amsfonts}
\usepackage{amsmath}
\usepackage{amssymb}
\usepackage{graphicx}
\usepackage{epsfig}
\usepackage{citesort}

\usepackage{dcolumn}
\usepackage{bm}

\begin{document}

\preprint{APS/XXX-XXA}

\title{Qubits entanglement dynamics modified by an effective atomic environment}
\author{I. Sainz}
\affiliation{Universidad de Guadalajara, Centro Universitario de los Lagos, Enrique Diaz de Leon, s/n, Lagos de Moreno, Jalisco, Mexico.}
\author{A. B. Klimov}
\affiliation{Departamento de F\'{\i}sica, Universidad de Guadalajara, Revoluci\'{o}n 1500, 44420 Guadalajara, Jalisco, Mexico.}
\author{Luis Roa}
\affiliation{Center for Quantum Optics and Quantum Information, Departamento de F\'{\i}sica,
Universidad de Concepci\'{o}n, Casilla 160-C, Concepci\'{o}n,
Chile.}

\date{\today}
\begin{abstract}
We study entanglement dynamics of a couple of two-level atoms resonantly interacting 
with a cavity mode and embedded in a dispersive atomic environment. We show that in the
absence of the environment the entanglement reaches its maximum value 
when only one exitation is involved. Then, we find
that the atomic environment modifies that entanglement dynamics
and induces a typical collapse-revival structure even for an initial 
one photon Fock state of the field.
\end{abstract}

\maketitle

\section{Introduction}

Entanglement generation between two atomic qubits has attracted considerable
attention during the last two decades due to its importance in various quantum
information processes \cite{Shor,Nielsen}. Those ideal processes, such as
quantum teleportation, quantum cryptography, and quantum computation
algorithms are strongly related to the capability of generating bipartite
entanglement \cite{Ekert,Deutsch,Bennett}. However, in real quantum systems
there are uncontrollable interactions with the surrounding environment which
usually lead to a decoherence resulting in the destruction of the
entanglement. Recently, several effects of different kinds of noisy environments,
specifically bosonic environment \cite{Yi,Bosonic,An,Kraus,Schneider,Benedict},
and fermionic environment \cite{Fermionic,Dawson,Lucamarini,Ma,Gao}
on the entanglement dynamics have been
extensively studied.
Especial effort was applied to find decoherence free entangled states
\cite{Lucamarini,Ma}. For instance, B. Kraus and J. I. Cirac \cite{Kraus} show
that two atoms can get entangled by interacting with a common source of
squeezed light and the steady state is maximally entangled even though the
modes are subjected to cavity losses. S. B. Zheng and G. C. Guo \cite{Zheng}
proposed a scheme to generate two-atom EPR states in such a way that the
cavity is only virtually excited. S. Schneider and G. J. Milburn \cite{Schneider}
show how the steady state of a dissipative
many-body system, driven far from equilibrium, may exhibit nonzero quantum
entanglement. Molmer and Sorensen have proposed a scheme for the generation
of multiparticle entangled states in ion traps without the control of the
ion motion.

Although the effect of the environment on the atomic entanglement is
usually destructive, in some specific situations two quantum systems can get
entangled in the process of their decaying to a common thermal bath \cite
{Braun,Benatti}. A similar effect was discussed in \cite{Plenio} where a
method of generation of entangled light from a noisy field has been proposed.
It was also shown \cite{Gao} that the interaction between two spins and an
itinerant electron environment leads to entanglement of the initially
unentangled spins.

In this article we study how an effective atomic environment modifies the
atomic entanglement generated in the course of resonant interaction of a
single mode of the cavity field with a couple of two-level atoms (the
so-called Dicke or Tavis-Cummings \cite{Tavis} model). Evolution of
entanglement in the two-atom Dicke model was previously studied in the case
of an ideal cavity in \cite{Tessier} and in the presence of a dissipative
environment in \cite{Tanas}. Our study is motivated by the following
physical situation: consider a cluster of two-level atoms  (resonant with a mode
of a cavity field) placed in a strong electric field (see e.g. \cite{Haroshe,Aoi}).
Physically it could be a cluster of polar moleculae.
The electric field generates a noticeable Stark shift so that most of
the atoms are detuned far from the resonance, except a very small portion of
them, whose dipole moments are approximately orthogonal to the field. Because the
atom-(quantum) field interaction times are much shorter than the typical
times of atomic diffusion, we can consider that the
orientation of the dipole moment is \textquotedblleft frozen\textquotedblright
and that the physical mechanism
of changing the atomic dipole orientation is a collision with the
cavity walls, since collisions between the atoms
in an atomic cluster are practically improbable.
In the process of interaction
with the cavity field the resonant atoms become entangled. We will study the simplest
situation where there are only two resonant atoms. Nevertheless, the
effect of non-resonant atoms on the dynamics of resonant ones is not
trivial. The dispersive interaction of the field mode with non-resonant
atoms leads to a modification of the field's phase which, in turn, affects
the evolution of resonant atoms. Thus, the non-resonant atoms play the role
of an effective dispersive environment whose whole effect could be expected
to reduce to a phase dumping \cite{Zurek}, and thus to the entanglement
decaying. Nevertheless, as it will be shown, the influence of such effective
environment is not always destructive but also leads to a constructive
interference, which reflects in, appearance of a system of collapses and
revivals of the atomic concurrence even in the presence of just a single
photon in a cavity.
The article is organized as follows: In section \ref{limit} we analytically
show, for some
specific initial conditions (non-excited atoms and the field in a Fock
state), that the entanglement of formation in a bipartite system of two-level
atoms interacting with a quantized mode reaches its maximum value when only
one excitation is involved and it decays as $1/n$ when $n\gg 1$,
being $n$ the number of photons in the initial Fock field state. In section
\ref{dickeAat} we derive the effective
Hamiltonian of noninteracting two-level (resonant) atoms and a cluster of $A$ atoms (far from resonance)
interacting simultaneously with a quantized mode and we find the evolution operator
when only one excitation is considered. In section \ref{EofF}
we study the effect of the dispersive atomic environment on the
entanglement dynamics generated by one excitation for two different initial conditions. In
section \ref{Conclusions} we summarize our results.

\section{Entanglement in the two-atoms Dicke model}   \label{limit}

By entanglement of two subsystems we mean the quantum mechanics feature whose
state can not be written as a mixed sum of products of the states of each
the the subsystems. In this case the entangled subsystems are no longer independent
even if they are spatially far separated. A measure, $E(|\phi\rangle)$, of
the degree of entanglement for a pure $|\phi\rangle$ state of a bipartite
system can be given by means of the entropy of von Neumann, of any of the two
subsystems. For a mixed state $\rho$ the entanglement of formation $E(\rho)$
between two bidimensional systems is defined as the infimum of the average
entanglement over all possible pure-state ensemble decompositions of $\rho$ 
\cite{Wootters}. Wootters found an analytic solution to this minimization
procedure in terms of the eigenvalues of the $R=\sqrt{\sqrt{\rho }\tilde{\rho}
\sqrt{\rho }}$ or $R'=\rho \tilde{\rho}$ non-Hermitian operators, where the
tilde denotes the spin flip of the quantum state. The solution for the
$\mathfrak{C}(\rho)$ concurrence associated with the entanglement of formation
of a mixed state of a bipartite of bidimensional subsystems is given by
$\mathfrak{C}(\rho )=\max \{0,\lambda _{1}-\lambda _{2}-\lambda _{3}-\lambda
_{4}\}$, where the $\lambda _{i}$'s are the square roots of eigenvalues of
the $R'$ operator and the eigenvalues of the $R$ operator, decreasingly ordered.
Throughout this article we consider this $\mathfrak{C}(\rho)$ as a measure of the
entanglement degree between the two resonance atoms $a$ and $b$.

\begin{figure} [t]
\includegraphics[angle=360,width=0.40\textwidth]{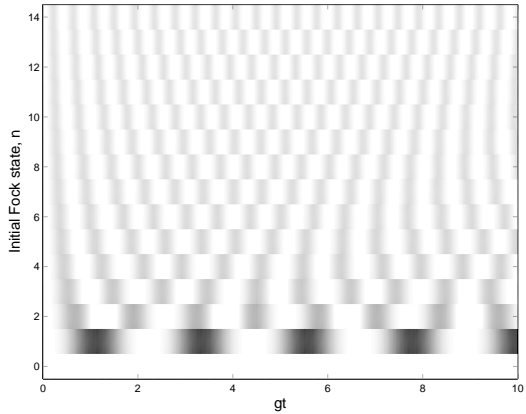}
\caption{Evolution ($gt$) of the concurrence
for initially unexcited $a$ and $b$
atoms and the field in a $n$ Fock state. Black means value $1$,
maximum entanglement, white means value zero, and greys mean
partial values of entanglement.}  \label{fig0}
\end{figure}

Let us consider two identical two-level atoms resonantly interacting with a
single-mode cavity field. The interaction Hamiltonian has the form 
\begin{equation}
H=g\{ a(s_{+a}+s_{+b})+a^{\dagger }(s_{-a}+s_{-b})\} ,
\end{equation}
where $s_{+j}=|1\rangle_{jj}\langle 0|$ and $s_{-j}=|0\rangle _{jj}\langle
1|$, with $|1\rangle _{j}$ and $|0\rangle _{j}$ being the excited
and ground eigenstates of $\sigma_{zj}$ of the $j$th atom ($j=a,b$), $a$ and $a^{\dagger }$ are,
respectively, the creation and annihilation operators for the cavity mode,
$g$ is the atom-cavity coupling strength. Considering initially a Fock field
state and both atoms in their ground states, the reduced atomic density
operator, at time $t$, is 
\begin{widetext}
\begin{equation}
\rho_{ab} =\frac{[nC_n(t)+n-1]^2}{(2n-1)^{2}}|0\rangle_a|0\rangle_{b\hspace{0.03in}a}\langle 0|_b\langle 0|
+\frac{nS_n^2(t)}{2n-1}|\psi^+\rangle _{ab\hspace{0.03in}ab}\langle\psi^{+}|
+\frac{n( n-1)[1-C_n(t)]^2}{( 2n-1)^{2}}|1\rangle
_{a}|1\rangle_{b\hspace{0.03in}a}\langle 1|_{b}\langle 1|, \label{rabb}
\end{equation}
\end{widetext}
where we have defined the functions:
\begin{eqnarray}
C_{n}(t)&=&\cos(\sqrt{2(2n-1)}gt), \nonumber \\
S_{n}(t)&=&\sin(\sqrt{2(2n-1)}gt), \nonumber
\end{eqnarray}
and the symmetric state:
\begin{equation}
|\psi^+\rangle_{ab}=(|0\rangle_a|1\rangle_b+|1\rangle_a|0\rangle_b)/\sqrt{2}.  \nonumber
\end{equation}
So, the $\mathfrak{C}(\rho_{ab})$ concurrence of the (\ref{rabb}) density operator is given
by
\begin{equation}
\mathfrak{C}(\rho_{ab})=\frac{nS_n^{2}(t)}{2n-1}
-2\frac{\sqrt{n(n-1)}}{(2n-1)^{2}}|nC_n(t)+n-1||1-C_n(t)|, \label{concurrence}
\end{equation}
when it is positive and is zero otherwise. Both terms on
the right side hand of Eq. (\ref{concurrence}) are zero for $n=0$ whereas
the second term is also zero for $n=1$ and, for other values of $n$, the
second term always reduce the concurrence. Therefore, as a function of the
number $n$ of excitations, the concurrence (\ref{concurrence}) acquires its
maximum value for $n=1$ at any time instant and it is given by $\mathfrak{C}(\rho_{ab})=\sin^2(\sqrt{2}gt)$. On the other hand, for $n\gg 1$
the concurrence (\ref{concurrence}) behaves as 
\begin{equation}
\lim_{n\gg 1}\mathfrak{C}(\rho_{ab})\approx\frac{1-\cos(\sqrt{2(2n-1)}gt)}{n}.
\end{equation}
The behavior of $\mathfrak{C}(\rho _{ab})\sim n^{-1}$ in the limit $n\gg 1$ was found
numerically by Tessier \textit{et al.} \cite{Tessier}.  Figure \ref{fig0} shows the concurrence as a function of the $n$ initial Fock state, and the $gt$ adimensional time. Black means value $1$, maximum entanglement, white means value zero, whereas greys mean partial values of entanglement. It can be seen that maximun value $1$ is only reached for the initial condition $|n=1\rangle$.
In the next section
we study how the concurrence is affected by the presence of an effective
atomic environment when only one excitation is involved.

\section{Effective Hamiltonian description}   \label{dickeAat}

We consider a collection of $A+2$ non-identical two-level atoms interacting
with a single mode of a quantized field in an ideal cavity. Two atoms,
labelled by subindexes $a$ and $b$, are resonant with the mode whereas the
other $A$ atoms interact dispersively with the mode. The Hamiltonian which
drives the unitary dynamics of the whole system under the rotating wave
approximation has the form 
\begin{eqnarray}
H&=&\omega_fa^{\dagger}a+\sum_{i=a,b}\omega_{i}s_{zi}+\sum_{j=1}^A\omega_{j}s_{zj}  \nonumber \\
&&+\sum_{j=a,b,1}^Ag_{j}(as_{+j}+a^{\dagger}s_{-j}) ,
\label{h}
\end{eqnarray}
where $a^{\dagger}$ and $a$ are the usual one mode field operators,
and $s_{z,\pm j}$ are the $z$ components of the Pauli
operators corresponding to the $j-$th two-level atom ($j=a,b,1,\ldots ,A$).
Atomic operators obey the standard $SU(2)$ commutation relations,
$[s_{+i},s_{-j}] =2s_{zi}\delta_{ij}$ and $[ s_{zi},s_{\pm j}]
=\pm s_{\pm i}\delta_{ij}$. Since the total number of excitations,
represented by the operator $\hat{N}=a^{\dagger }a+\sum_{j=a,b,1}^{A}s_{zj}$,
is an integral of motion, the above Hamiltonian can be rewritten as
follows: 
\begin{equation}
H=\omega_{f}\hat{N}+H_{int},
\end{equation}
with 
\begin{eqnarray}
H_{int} &=&\sum_{j=1}^A\Delta _{j}s_{zj}+
\sum_{i=a,b}g_{i}(as_{+i}+a^{\dagger}s_{-i})  \nonumber \\
&&+\sum_{j=1}^Ag_{j}(as_{+j}+a^{\dagger}s_{-j}),
\label{hint}
\end{eqnarray}
where $\Delta_{k}=\omega_{k}-\omega_{f}$, $k=1,\ldots ,A$ are the
detunings between the transition of the $k$th atom and the mode frequency.
Now, we assume that all $A$ atoms are far from the resonance, so that
$\Delta_{j}\gg g_{j}$, $j=1,2,\ldots ,A$. The effective Hamiltonian,
approximately describing the interaction process, can be obtained from the
interaction Hamiltonian (\ref{hint}) by using the method of Lie rotations 
\cite{Small,Sainz}, namely by applying to the Hamiltonian (\ref{h}) the
following unitary transformation 
\begin{equation}
V=e^{\hat{B}},\hspace{0.2in}\hat{B}=\sum_{j=1}^{A}\epsilon_{j}(as_{+j}-a^{\dagger}s_{-j}),
\end{equation}
where $\epsilon_{j}=g_{j}/\Delta_{j}\ll 1$. Neglecting terms of order
higher than $\epsilon _{j}^{2}$, we obtain the following effective
Hamiltonian: 
\begin{widetext}
\begin{equation}
H_{eff}=\sum_{j=1}^A(\Delta_{j}+g_j\epsilon_j(1+2a^{\dagger}a))s_{zj}+\hat{\lambda}
\sum_{i=a,b}g_i(as_{+i}+a^{\dagger }s_{-i})+\frac{1}{2}
\sum_{j,i=1}^{A}g_{i}\epsilon_j(s_{-i}s_{+j}+s_{+i}s_{-j}),  \label{heff1}
\end{equation}
\end{widetext}
where we have defined the operator $\hat{\lambda}=1+\sum_{j=1}^{A}\epsilon_{j}^{2}s_{zj}$.

The last term in (\ref{heff1}) represents an effective dipolar interaction
between the non-resonant atoms, and its contribution to the system dynamics
strongly depends on the internal resonance condition between atomic
frequencies. Let us consider randomly distributed frequencies, such that
they satisfy the condition $\Delta _{j}-\Delta _{i}\gg g$,
$i,j=1,2,..A$. Then, the terms $i\neq j$ in the last sum of the effective
Hamiltonian (\ref{heff1}) rapidly oscillate and can be neglected. Finally,
the effective Hamiltonian, up to a constant energy shift, becomes: 
\begin{eqnarray}
H_{eff}&=&\sum_{j=1}^A\Delta_js_{zj}+(1+2a^{\dagger}a)
\sum_{j=1}^Ag_{j}\epsilon_js_{zj}  \nonumber \\
&&+\hat{\lambda}\sum_{i=a,b}g_i(as_{+i}+a^{\dagger }s_{-i}).
\nonumber
\end{eqnarray}
In the given approximation the total number of excitations \textquotedblleft
stored\textquotedblright in the non-resonant atoms is a constant of motion,
which reflects a dispersive character of interaction. The first term in the
above equation represents just transition frequency shifts of the non-resonant
atoms, and commutes with the rest of the terms (so, it can be
taken out of the Hamiltonian). The second term is the dynamic Stark shift
and its contribution to the resonant dynamics, described by the last term,
strongly depends on the state of the non-resonant atoms, which can be
considered as a kind of atomic environment.

Since the maximum entanglement in the system of two resonant atoms is
reached when the total number of excitation is one, $N=1$, we consider
exclusively this situation. So, under the constraint that there is only one
photon, the corresponding evolution operator can be found and, in the standard
tensor product basis, it is given by 
\begin{widetext}
\begin{equation}
U(t)=\left[
\begin{array}{cccc}
e^{i\hat{y}t/2}\hat{A}_{n+1} & -ig\hat{\lambda}a \hat{L}_n e^{i\hat{y}t/2} & -ig\hat{\lambda}a\hat{L}_n e^{i\hat{y}t/2} & 0 \\
-ig\hat{\lambda}a^{\dagger } \hat{L}_{n+1} e^{i\hat{y}t/2} & \hat{Y}_n & \hat{Y}_n-1 & -ig\hat{\lambda}a\hat{L}_{n}e^{-i\hat{y}t/2} \\
-ig\hat{\lambda}a^{\dagger }\hat{L}_{n+1}e^{i\hat{y}t/2} & \hat{Y}_n-1 & \hat{Y}_n & -ig\hat{\lambda}a\hat{L}_{n}e^{-i\hat{y}t/2} \\
0 & -ig\hat{\lambda}a^{\dagger }\hat{L}_{n+1}e^{-i\hat{y}t/2} & -ig\hat{\lambda}a^{\dagger }\hat{L}_{n+1}e^{-i\hat{y}t/2} & e^{-i\hat{y}t/2}\hat{A}_{n}^{\ast }
\end{array}
\right] ,  \label{EO}
\end{equation}
\end{widetext}
where we have defined the operators: 
\begin{eqnarray}
\hat{L}_{n} &=&\frac{\sin \hat{\Omega}_{n}t}{\hat{\Omega}_{n}} \nonumber \\
\hat{Y}_{n} &=&\frac{1}{2}( e^{i\hat{y}t/2}\hat{A}_{n}^{\ast }+e^{-i
\hat{y}t/2}\hat{A}_{n+1})  \nonumber \\
\hat{A}_{n} &=&\cos \hat{\Omega}_{n}t+i\frac{\hat{y}}{2}\hat{L}_{n},  \nonumber
\end{eqnarray}
and the Rabi frequencies $\hat{\Omega}_{n}$ depend on the field and on the
environment variables as follows: 
\begin{equation}
\hat{\Omega}_{n}=\left[ \left( \frac{\hat{y}}{2}\right) ^{2}+2g^{2}\hat{
\lambda}^{2}\hat{n}\right] ^{1/2},\quad \frac{\hat{y}}{2}
=\sum_{j=1}^{A}g_{j}\epsilon _{j}s_{zj}.  \nonumber
\end{equation}
So, the dynamics depends on the distribution of the different Rabi
frequencies which appear as a contribution of $A$ non-resonance
distinguishable two-level atoms and it also depends on the initial state of
the whole system.

\section{Evolution of the Entanglement}   \label{EofF}

Now, let us suppose that the environment atoms are prepared in a coherent
superposition of excited and ground states: 
\begin{widetext}
\begin{equation}
|\Psi_{0}\rangle_{env}=\prod{j=1}^{A}\frac{|0\rangle_{j}+|1\rangle_{j}}{\sqrt{2}}
\frac{1}{2^{A/2}}\sum_{s_{1}}\sum_{s_{2}}\ldots\sum_{s_{A}}|s_{1}+1/2\rangle_{1}
|s_{2}+1/2\rangle_{2}\ldots|s_{A}+1/2\rangle_{A},
\label{envst}
\end{equation}
\end{widetext}
where the last sum is taken over all possible binary vectors
$\vec{s}=\left\{ s_{1},s_{2},\ldots s_{A}\right\}$, $s_{j}=s_{zj}=\pm 1/2$.
We capture the key features of the concurrence evolution by
considering two particular cases of initial conditions. First, we suppose
that the resonance atoms are initially in the ground state and that the field is
in the one photon Fock state: 
\begin{equation}
|\Psi\rangle =|0\rangle_{a}|0\rangle_{b}|1\rangle_{f}|\Psi _{0}\rangle_{env}.  \label{ini1}
\end{equation}
Applying the evolution operator (\ref{EO}) to the (\ref{ini1}) state and
tracing up over the field and the off-resonance atomic environment, we
obtain for the resonant atoms the following reduced density operator: 
\begin{widetext}
\begin{equation}
\rho_{ab}=\frac{1}{2^A}\sum_{m=1}^{2^{A}}\{\sin ^{2}(\Omega_{1}^{(m)}t)
|\psi^+\rangle_{ab\hspace{0.03in}ab}\langle \psi^+|+\cos ^{2}( \Omega _{1}^{(m)}t)|0\rangle_{a}|0\rangle_{b\hspace{0.03in}a}\langle 0|
_{b}\langle 0|\},  \label{dens1}
\end{equation}
\end{widetext}
with 
\begin{eqnarray}
\Omega_1^{(m)}&=&\frac{\tilde{g}^2}{4}(\overrightarrow{\epsilon}
\cdot\vec{s}^{(m)})^2=\frac{\tilde{g}^2}{4}(
\sum_{j=1}^A\epsilon_js_j^{(m)})^2,  \label{q} \\
\lambda_m &=&1+(\overrightarrow{\epsilon}^2\cdot\vec{s}^{(m)})
=1+\sum_{j=1}^A\epsilon_j^2s_j^{(m)},
\end{eqnarray}
where $\vec{s}^{(m)}$ is a certain arrangement of the $\vec{s}$ vector and
$s_j^{(m)}=\pm 1/2$. We have neglected order corrections higher than or
equal to $\epsilon$ on the amplitudes. From now on the coupling constants
for field-environment atoms are taken to be equal, $g_j=\tilde{g}$ for all 
$j$. 
\begin{figure}[t]
\includegraphics[angle=360,width=0.40\textwidth]{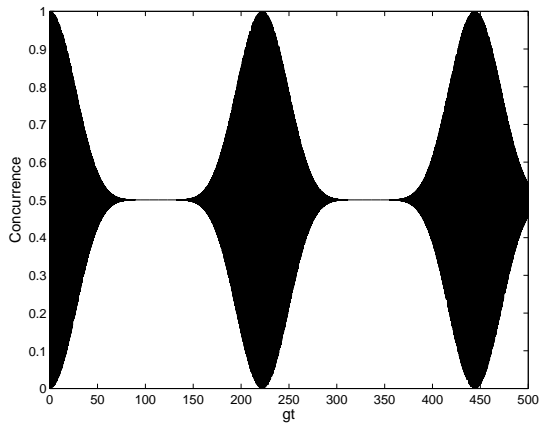}
\caption{Evolution of the concurrence for initially unexcited $a$ and $b$
atoms and the field in the one-photon Fock state, with $A=7$,
$\tilde{g}/g=1$, and $\tilde{g}/\bar{\Delta} =0.1$.}  \label{fig1}
\end{figure}
The concurrence corresponding to the density matrix
(\ref{dens1}) takes the form 
\begin{equation}
\mathfrak{C}(\rho_{ab})=\frac{1}{2^A}\sum_{m=1}^{2^A}\sin^{2}(\Omega_1^{(m)}t).  \label{c1}
\end{equation}

If the resonant atoms are initially prepared in the symmetric one excitation
state and the field is in the vacuum Fock state, then the initial state of
the whole system is the following tensor product: 
\begin{equation}
|\Psi\rangle=\frac{|0\rangle_a|1\rangle_b+|1\rangle_a|0\rangle_b}{\sqrt{2}}
|0\rangle_f|\Psi_0\rangle_{env}.
\end{equation}
In a similar way as described in the previous case (\ref{ini1}) the
concurrence takes the form 
\begin{equation}
\mathfrak{C}(\rho_{ab})=\frac{1}{2^A}\sum_{m=1}^{2^A}\cos^2(\Omega_1^{(m)}t).  \label{c2}
\end{equation}
It is worth noting that the concurrences (\ref{c1},\ref{c2}) are composed of
many Rabi frequencies, which leads to a structure similar to the collapses and
revivals in the Jaynes-Cummings model \cite{Tavis,Banacloche}. Nevertheless, in the
present case the set of different frequencies is due to the presence of the
dispersive atomic environment in contrast to the standard JCM where
different Rabi frequencies appear as contributions of different Fock field states
(recall that a well-defined collapse-revival structure requires a
significant number of excitations \cite{Banacloche,Retamal}).

Let us consider randomly distributed numbers $\Delta_{j}$, $j=1,\ldots,A$
with the mean $\bar{\Delta}$ and the standard deviation $\sigma_{\Delta}$.
Then, the numbers $\epsilon_{j}$, $\epsilon _{j}^{2}$ have the following
mean values and standard deviations: 
\begin{eqnarray}
\bar{\epsilon} &\approx &\frac{\tilde{g}}{\bar{\Delta}}\left( 1+\frac{\sigma
_{\Delta }^{2}}{\bar{\Delta}^{2}}\right) ,\hspace{0.33in}\sigma _{\epsilon
}\approx\frac{\tilde{g}\sigma _{\Delta }}{\bar{\Delta}^{2}}, \\
\overline{\epsilon ^{2}} &\approx &\frac{\tilde{g}^{2}}{\bar{\Delta}^{2}}
\left( 1+\frac{3\sigma _{\Delta }^{2}}{\bar{\Delta}^{2}}\right) ,\hspace{
0.15in}\sigma _{\epsilon ^{2}}\approx\frac{2\tilde{g}^{2}\sigma _{\Delta }}{ 
\bar{\Delta}^{3}}.
\end{eqnarray}
First, we will find the distribution of the (\ref{q}) quantities. We assume
that $\sigma_{\epsilon}\ll\bar{\epsilon}$ and $\sigma_{\epsilon^{2}}\ll 
\overline{\epsilon^{2}}$. Then, there are $A+1$ peaks corresponding to
different values of the number of positive components of $\vec{s}$,
$k=0,\ldots A$; for a given value of $k$ there are $C_{k}^{A}=A!/[
k!(A-k)!]$ values of $(\vec{\epsilon}\cdot\vec{s})$ and
$(\overrightarrow{\epsilon^{2}}\cdot \vec{s})$
which are normally distributed in accordance with the central limit theorem.
For the $k$th peak, the mean value and the standard deviation are given by 
\begin{eqnarray}
\langle( \vec{\epsilon}\cdot \vec{s})\rangle_{k}
&=&(k-A/2)\bar{\epsilon},\hspace{0.24in}\sigma_{\epsilon,k}
=\sigma_{\epsilon}\sqrt{\frac{k(A-k)}{A-1}},  \nonumber \\
\langle(\overrightarrow{\epsilon^2}\cdot\vec{s})\rangle_{k}
&=&(k-A/2)\overline{\epsilon^{2}},\hspace{0.12in}
\sigma_{\epsilon^2,k}=\sigma_{\epsilon^{2}}\sqrt{\frac{k(A-k)}{A-1}}. \nonumber
\end{eqnarray}
Note that the first and the last peaks are infinitely narrow.

The Rabi frequency distribution has $A+1$ peaks (now the summation in (\ref{c1}, \ref{c2})
is from $k=0$ to $A$), and the frequency corresponding to
the $k$th peak can be approximated as follows: 
\begin{equation}
\Omega_{k}\approx\sqrt{2}g[1+\frac{\tilde{g}^{2}}{\bar{\Delta}^{2}}
( k-\frac{A}{2})(1+\frac{\tilde{g}^{2}}{4g^{2}}(k-\frac{A}{2}))].  \nonumber
\end{equation}
Thus, the expressions (\ref{c1}) and (\ref{c2}) for the concurrence can be
approximated as follows: 
\begin{equation}
\mathfrak{C}(\rho_{ab})\approx \frac{1}{2}\left[1\pm \frac{1}{2^A}\text{Re} \sum_{k=0}^A
C_k^A\exp( 2i\Omega_k t) \right].  \label{Cc}
\end{equation}
The $d_{k}$ separation between the $k$th and the $(k+1)$th
peaks, and the $\delta_{k}$ width of the $k$th peak are 
\begin{eqnarray}
d_{k} &=&\sqrt{2}g[\frac{\tilde{g}^{2}}{\bar{\Delta}^{2}}+
\frac{\tilde{g}^{4}}{4g^{2}\bar{\Delta}^{2}}(2k-A+1)],  \nonumber \\
\delta_{k} &=&\sqrt{2}g[1+(k-\frac{A}{2})+\frac{\tilde{g}
^{2}}{4g^{2}}(k-\frac{A}{2})^{2}] \sigma_{\epsilon^{2},k}.  \nonumber
\end{eqnarray}
Then, considering the approximation of narrow peaks, $\delta_{k}\ll d_{k}$,
i.e. $\sigma_{\Delta}\ll \bar{\Delta}/\sqrt{A}$, the sum in (\ref{Cc}) can
be represented as a sum of Gaussians, that is 
\begin{equation}
\mathfrak{C}(\rho_{ab})\approx \frac{1}{2}\pm \frac{\bar{\Delta}^{2}e^{2\sqrt{2}igt} }{2
\tilde{g}^{2}\sqrt{2A}\sigma }\text{Re}\sum_{k=-\infty }^{\infty }e^{-\frac{
\left( gt-\frac{\pi k\bar{\Delta}^{2}}{\sqrt{2}\tilde{g} ^{2}}\right) ^{2}}{
2\sigma ^{2}}} ,  \label{c22}
\end{equation}
with the width 
\begin{equation}
\sigma ^{2}=\frac{\bar{\Delta}^{4}}{2\tilde{g}^{4}A}-it\frac{\sqrt{2}\bar{
\Delta}^{2}}{8g},  \label{sigma}
\end{equation}
which grows with time. The (\ref{c22}) sum reveals the collapse-revival
structure of the concurrences (\ref{c1}) and (\ref{c2}). The first collapse
happens when 
\begin{equation}
gt_{c}\sim \frac{\bar{\Delta}^{2}}{\sqrt{A}\tilde{g}^{2}},  \label{tcol}
\end{equation}
and it is followed by revival at time: 
\begin{equation}
gt_{R}\sim \pi k\frac{\bar{\Delta}^{2}}{\sqrt{2}\tilde{g}^{2}}, \hspace{0.2in} k=1,2,.... \label{trev}
\end{equation}
In Figure \ref{fig1} we show the exact evolution of the concurrence for
initially unexcited atoms and the field in the one-photon Fock state in the presence of
the environment atoms. One can observe that the entanglement also
reaches its maximum value.
We can estimate from (\ref{tcol}) and (\ref{trev}) the time scale required to observe the environment
induced collapse-revival structure. Taking the typical values of the interaction constant from \cite{Haroshe}:
$g/2\pi=24 kHz$ and $\bar{\Delta}/2\pi \sim 70 kHz$, we obtain $t_R\sim 100 \mu s$ which is of order of the passage time of the atom through the cavity  (cold atoms, $v\sim 100 m/s$) and less than the photon lifetime $\sim 160 \mu s$. The collapse time is $\sqrt{A}$ times less than $t_R$.

\section{Conclusions}   \label{Conclusions}

In summary, we have studied the dynamics of the
concurrence of two atoms resonantly interacting with a cavity mode in the
presence of many off-resonance atoms. We have shown that, for random
distribution of atomic detunings and initially symmetric excitation of
non-resonant atoms, the coherent influence of the environment can be
separated from the dephasing. The coherence influence of the environment
reflects in the appearance of the collapse-revival structure
of the concurrence, with an average value of
one half. This behavior is induced only by the presence of the dispersive
environment. It is worth noting that the entanglement of formation in
revival periods has as extreme value $1$, which means that the
state of the bipartite system becomes pure at those times.

\begin{acknowledgments}
This work was partially supported by Grants FONDECyT No. 1030671, Milenio
ICM P02-49F. The work of A. B. Klimov is partially supported by grant
PROMEP/103.5/04/1911.

The authors thank Carlos Saavedra and Jose Aguirre for valuble dicussions.
\end{acknowledgments}

\end{document}